# Vertical, electrolyte-gated organic transistors: continuous operation in the MA/cm² regime and use as low-power artificial synapses

*Jakob Lenz[1], Fabio del Giudice[1,#], Fabian R. Geisenhof[1], Felix Winterer[1], R. Thomas Weitz[1,2,3]\**

1. Physics of Nanosystems, Department of Physics, Ludwig-Maximilians-Universität München, Amalienstrasse 54, 80799 Munich, Germany
2. Nanosystems Initiative Munich (NIM), Schellingstrasse 4, 80799 Munich, Germany
3. Center for NanoScience (CeNS), Ludwig-Maximilians-Universität München, Geschwister-Scholl-Platz 1, 80539 Munich, Germany
# current address: Walter-Schottky Institute, Technical University Munich, Am Coulombwall 4, 85748 Garching, Germany
* email: thomas.weitz@lmu.de

**Abstract:** Organic semiconductors are usually not thought to show outstanding performance in highly-integrated, sub 100 nm transistors. Consequently, single-crystalline materials such as SWCNTs, $MoS_2$ or inorganic semiconductors are the material of choice at these nanoscopic dimensions. Here, we show that using a novel vertical field-effect transistor design with a channel length of only 40 nm and a footprint of 2 x 80 x 80 nm², high electrical performance with organic polymers can be realized when using electrolyte gating. Our organic transistors combine high on-state current densities of above 3 MA/cm², on/off current modulation ratios of up to $10^8$ and large transconductances of up to 5000 S/m. Given the high on-state currents at yet large on/off ratios, our novel structures also show promise for use in artificial neural networks, where they could operate as memristive devices with sub 100 fJ energy usage.

Organic semiconductors are promising components for novel flexible electronics such as displays and sensors. Large-scale processability via printing and the inherent flexibility of organic materials are two key advantages in this context. Attributes that do usually not come to mind when discussing organic materials are high current densities in the MA/cm², large transconductances in the 10 – 100 S/m regime, low power operation or low supply voltages in the sub-Volt regime as they are common in highly integrated nanoscale transistors[1]. More specifically, while for example state-of-the-art single-walled carbon nanotube (SWCNT) or $MoS_2$ field-effect transistors (FETs) are able to sustain current densities above MA/cm² [1–3], organic transistors are currently only able to operate at tens of kA/cm² [4,5]. While such high current densities in the MA/cm² are crucial for realizing highly integrated, high-performance electronics[1], low power operation on the other hand is critical for the operation in handheld devices or in neuronal networks. In the latter, the energy per switching event is required to be reduced down to the sub-pJ regime, implying the need for high on-state conductance, large current modulation ratios and low voltage operation.

Up to now, organic materials have not been able to meet these diverse demands and can either drive high currents (e.g. SWCNTs[3]) or have a large on/off ratio (most organic semiconductors[6]), but not both. Here, we discuss a novel nanoscopic device design that is based on a vertical transistor structure and enables electrolyte-gated organic semiconductors to drive MA/cm² currents combined with on/off ratios of $10^8$. The small channel length of below 50 nm and nanoscopic device footprint of 2 x 80 x 80 nm² allows their use in highly integrated circuits. Additionally, we show that our novel device design makes organic semiconductors prime candidates for use in low power neuromorphic computing.

The established approach to realize nanosopic channels is the planar transistor geometry which we have also started our investigations with. Gold-Palladium contacts with a separation between 30 nm and 500 nm have been patterned using conventional electron-beam lithography, metal evaporation and lift-off. We have used a diketopyrrolopyrrole–terthiophene donor–acceptor polymer (PDPP, Fig. 1a inset) as prototype semiconductor for our device structures[7]. A PDPP solution was deposited via doctor blading yielding a sub-monolayer thin film and after deposition of the ionic liquid 1-ethyl-3-methylimidazolium bis(trifluoromethylsulfonyl)imide ([EMIM][TFSI])[8] the transistors were electrically characterized in vacuum or ambient atmosphere. Electrolyte gating has been established in the past as reliable method to enable large gate coupling[9–12]. Consequently, short channel effects in lateral sub-micrometer devices have been drastically reduced and is also the reason why we utilize it here[13]. A typical transfer curve of a transistor with 220 nm and 40 nm channel length with corresponding SEM images are shown in Fig. 1a-d. As common for electrolyte-gated transistors, due to the large capacitance of the electrolyte (11 µF/cm² in our case[8]), the transistors can be operated at gate-source voltages $V_{GS}$ below 2 V. The respective output characteristics can be found in Supplementary Fig. 1. Our transistors do not show any signs of contact resistance, and high on-off current ratios of $10^5$ are observed. Furthermore, our devices show large transconductances of up to 6 S/m. Even though these values are lower than what has been shown before e.g. in P3HT transistors[14], the electrical characteristics are encouraging, especially given the fact that the semiconducting film is only 2 nm thick (Fig. 1d). Finally, our on-state resistance of $\rho_{on,PDPP}$ = 5.4x10$^{-4}$ Ωm compares favorably to P(NDI2OD-T2) transistors ($\rho_{on,P(NDI2OT-T2)}$ = 0.12 Ωm)[15], the only other monolayer polymer transistors we are aware of.

While the electrical characteristics of these lateral transistors are very encouraging, for highly integrated applications a smaller footprint of the channel is needed. Therefore, we have realized transistors with vertical contact separation that can be gated with an electrolyte. Vertical organic field-effect transistors (VOFETs) have been established as promising approach to achieve nanoscopic source-drain contact separation and a nanoscopic footprint without the need for high-resolution pattering[16,17]. In the past years, several groups have reported high device performance of VOFETs[18–26]. We have now introduced electrolyte gating to VOFETs in a novel device structure and found surprisingly improved performance compared to the above discussed lateral device design as well as to existing VOFETs. The device fabrication is shown in Fig. 2a-f. Two gold electrodes serving as source and drain are patterned for convenience by electron beam lithography. The channel length $L_c$ of the transistors can be controlled via the thickness of the insulating SiO$_2$ layer which was varied between 35 and 50 nm in our experiment, but can in principle be chosen arbitrarily and does not depend on the use of e-beam lithography. The SiO$_2$ and titanium layers are subsequently removed across a width d$_c$ with 1% HF acid resulting in an underetched top contact. The magnitude of d$_c$ can be controlled via the etching time and a smaller d$_c$ enables better control of the channel with the electrolyte, since the ions have to diffuse a smaller length to control the entire channel. The channel area A$_{ch}$ (i.e. its footprint) is given by A$_{ch}$ = 2 x w$_{bel}$ x d$_c$, while w$_{bel}$ is the width of the bottom electrode. In our studies we tested devices with d$_c$ between 80 and 120 nm and w$_{bel}$ between 80 nm and 500 µm. Subsequent semiconductor deposition and reactive ion etch (RIE) finishes the transistors. The directional etching process allows us to remove the semiconductor everywhere except below the top electrode, since the latter serves as etching mask for the semiconductor (see Fig. 2g)[27]. A cross sectional SEM image of a device broken through the transistor channel is depicted in Fig. 2h. For the electrical measurement we again utilize the liquid electrolyte [EMIM][TFSI][8] to control the charge carrier density in the semiconducting channel. In this device architecture, the electrolyte ions diffuse sideways directly into the vertical bulk channel. Typical I-V characteristics of an electrolyte gated VOFET with a channel area of A$_{ch}$ = 1.6 x 10$^{-11}$ m$^2$, a channel length of L$_c$ = 40 nm and a bottom electrode width of w$_{bel}$ = 100 µm are shown in Fig. 3a and b. As can be seen in the output characteristics in Fig. 3a, the transistor exhibits good saturation behavior, which is indispensable for e.g. AMOLED displays and all other applications, where transistors are used as current source[28,29]. Also, it is good proof that even though the channels are only 40 nm short, the high gate coupling via the ionic liquid fully controls the charge carrier density in the channel. The corresponding transfer curves are shown in Fig. 3b where one can clearly see the large on/off ratio of up to 10$^8$. The drain-source voltage (V$_{DS}$) dependent shift of the threshold voltage stems from voltage-induced drain barrier lowering[30] and is the only short-channel effect we have observed. The current density in Fig. 3a and B is J$_{-0.3V}$ = 32.9 kA/cm$^2$ at V$_{DS}$ = -0.3 V and J$_{-10mV}$ = 11.3 kA/cm$^2$ at V$_{DS}$ = -10 mV. Additionally, these vertical transistors reveal an outstanding transconductance of g$_m$ = 58.5 S/m (see Supplementary Fig. 2a). Currently, we can only speculate about the microscopic reason for the significantly improved electrical characteristics of the vertical compared to lateral device designs. We anticipate that it is related to a favorable morphology of the semiconductor and a better charge injection.

The here developed geometry cannot only be used with PDPP polymers, but provides a general platform for nanoscale organic high-current transistors. For example, I-V characteristics for a P3HT transistor ($A_{ch}$ = $10^{-12}$ m$^{-2}$), fabricated in the same manner as described in Fig. 2 with a comparable current density ($J_{-0.3V}$ = 129 kA/cm$^2$) and on-off-ratio ($10^6$) can be found in Supplementary Fig. S3.

The combination of short channels with electrolyte gating is of interest for a number of potential applications where low-voltages, high on-state current densities or large transconductances are required. Examples are the driving of OLEDs[16], sensors[14] or memristors[31–35]. We first focus on explaining charge transport at high current densities in detail before we show use of the electrolyte gated VOFETs in the field of low-power memristive devices.

### Operation of organic transistors in the MA/cm² regime

In the transistor discussed in Fig. 3a and b, the total resistance $R_{tot}$ in the on-state with a maximum on current of $I_{on}$ = 5.6 mA at $V_{DS}$ = -0.3 V is $R_{tot}$ = 54 Ω. Control experiments revealed similar total resistances for only the measurement setup including contact resistances. Apparently, the channel resistance seems to be almost negligible in this specific device geometry and $I_{on}$ is predominantly limited by contact and lead resistances. To find an upper limit for $I_{on}$, we have increased the relative resistance of the channel by reducing the channel area $A_{ch}$ to a nanoscopic 2 x 80 x 80 nm² (the resistance of the current leads stays the same compared with the previously described devices, note that the channel length $L_c$ was left unchanged). The maximum current density for these nanoscopic transistors is $J_{-0.3V}$ = 2.7 MA/cm$^2$ (on-off ratio $10^7$) and $J_{-10mV}$ = 89.9 kA/cm$^2$, at $V_{DS}$ = -0.3 V and -10 mV respectively (see Fig. 4a and b).

Since these current densities are exceptionally large for organic transistors (see below for a detailed comparison to other state-of-the-art transistors), we have made a series of test experiments to make sure that the current is really flowing through the organic semiconductor and is not caused by shorts in the device. First, we have found that a transistor structure prior to semiconductor deposition only showed insulating behavior and no response to ionic liquid gating (Supplementary Fig. 4a). Second, after removing the ionic liquid and depositing the semiconductor, the same transistor showed a clear response to the ionic liquid gate with a maximum current density of about 3 MA/cm² (Supplementary Fig. 4b). Finally, we purposely removed the organic semiconductor of the previously working VOFET by 10 minutes sonication in a xylene bath followed by an anisotropic $O_2$ plasma, and again observed insulating behavior (Supplementary Fig. 4c). Furthermore, SEM images of once working VOFETs after electrical measurements revealed no damage of the gold electrodes. In fact, the failure current density of gold nanowires exceeds the maximum observed current density in our electrolyte gated VOFETs by at least a factor of 100 [36]. Additionally, we have also tested if we can intentionally destroy our device by applying a higher drain-source voltage. Indeed, at $V_{DS}$ = -1.8 V, the current decreases slowly down to the noise level (Supplementary Fig. 5b). Since this voltage is close to the electrochemical window reported for [EMIM]/[TFSI] at room temperature[37], the failure mechanism might be attributed to an electrochemical reaction of the organic semiconductor with the electrolyte. The combined experimental evidence is a clear proof, that the current is carried exclusively by the organic semiconductor. Additionally, we have assessed the cycling stability of the VEOFETs and measurements of 20 $I_D$-$V_{GS}$ cycles showed no obvious drift in the threshold voltage or decrease of $I_D$ (see Supplementary Fig. 6). Finally, we have measured the same transistor after three months of storage in a desiccator which showed the same on current and only a small shift of $V_{th}$ (see Supplementary Fig. 7).

For possible applications of the novel VOFETs, it is also important to test for how long the devices can sustain the ~ 1 MA/cm² currents. This question is especially critical, since previous VOFETs showed significant degradation due to Joule heating if operated for more than few ms at current densities above 1 kA/cm², and are therefore only operated in pulsed mode[5]. In contrast, as shown in Fig. 4c, our transistors can be operated continuously for at least 50 min at MA/cm² current densities without significant degradation of the current (see also Supplementary Fig. 8). We assume, that the stability at these high current densities stems from the favorable device geometry, since the small channel width and length provide intimate contact of the semiconductor to the source and drain contacts as well as to the $SiO_2$, which act as heat sink that allow for a rapid dissipation of the developing heat. Additional heat might be dissipated by the ionic liquid that has entirely penetrated the semiconductor. This is also probably the reason that neither a self-heating induced N-shaped negative differential resistance reported for inorganic transistors[38] nor a S-shaped negative differential resistance recently presented for organic permeable-base transistors[39] was found in our devices.

The observed current densities are uniquely high for organic transistors and are enabled by the short contact distances, efficient gating and the excellent electrical performance of the used organic polymers. Fig. 4d summarizes and compares the device performances of several state-of-the-art VOFETs and lateral FETs with respect to their on-state current density and on-off ratio. The performance of our electrolyte gated VOFETs exceeds the best vertical organic transistors and is in fact comparable to inorganic vertical transistors based on e.g. GaAs[40,41]. This is particularly surprising, since our VOFETs were operated only at drain-source voltage of -0.3 V and -10 mV, respectively, which is at least a factor 4 smaller than the operation voltage of inorganic vertical FETs. Furthermore, our VOFETs also perform well compared to SWCNT, $MoS_2$ and FIN-FET[42] devices. Additionally, the outstanding properties are not limited to PDPP and are comparable for different polymers, e.g. P3HT, which demonstrates that this device architecture can be expected to be suitable for a wide range of semiconductors. Finally, our transistors show large transconductances of above 5000 S/m (see Supplementary Fig. 2b) – larger than for example SWCNT network FETs[3] or PEDOT:PSS FETs[14,43]. A more detailed comparison of transconductances is given in Table T1.

### VOFET based low-power memristive devices

The large on-state conductances, high on-off ratios and low $V_{GS}$ operational voltages make our novel device design also suitable for ultra-low power electronics. For example, we can operate our devices at a $V_{DS}$ of only 10 µV and still obtain on/off ratios of $10^2$ (see Supplementary Fig. 9). Such low power operation is especially relevant for applications in artificial neural networks as memristive devices. More specifically, our VOFETs combine the ability for low voltage operation with a small footprint, large on/off ratio, high switching speed, long-term stability of the electrical performance and the use of electrolyte gating[31]. To prove the general usability of our devices in this field, we show artificial synaptic behavior with short- and long-term plasticity (STP and LTP). As previously reported for electrolyte gated OFETs, the contact to the liquid electrolyte can be seen as the presynaptic- and the source electrode as the postsynaptic terminal[43,44]. Upon a voltage pulse at the gate electrode (corresponding to a presynaptic potential spike), the increase in $I_D$ can be viewed as the excitatory post-synaptic current (EPSC), which represents the synaptic strength. Before a presynaptic spike, the anions and cations are randomly distributed in the liquid electrolyte. A short negative voltage pulse causes anions to penetrate into the bulk of the semiconductor, leading to an accumulation of free holes in the semiconducting channel. These charge carriers contribute to the EPSC upon an applied source drain voltage. After the presynaptic spike there is no driving force for the ions to remain in the semiconductor, hence they slowly return to a random distribution and the EPSC decays. The EPSC change over time is regarded as synaptic plasticity that can be distinguished in STP and LTP. While STP is more important for application of memristive elements in computational applications, LTP is more important in learning[31]. As we show below, with our novel device geometry we can tune the relative strength of STP and LTP via the device design, thus making the layout suitable for a wide range of potential applications. Paired-pulse facilitation (PPF) is a possibility to simulate STP[44]. Fig. 5a shows the EPSC where the amplitude of the second postsynaptic response $A_2$ = 608 µA is amplified compared to the first one $A_1$ =10.5 µA by a factor of 58. Since before the second presynaptic spike the ions have not returned to a complete random distribution, these residual ions contribute to the second presynaptic spike resulting in an increased EPSC.

For long-term memory formation it is necessary to transform STP to LTP. LTP in our electrolyte gated VOFETs is shown in Fig. 5b. After six pulses (-0.8 V, 50 ms) with an inter-spike interval of 2.5 s, an increase of the EPSC after each pulse and an obvious nonvolatile channel current is measured, which constitutes memory formation. Another method to realize LTP is by increasing the magnitude of the gate pulse (Supplementary Fig. 10). For applications that rely on LTP, storage of the state for more than $10^3$ s would be favorable. The magnitude of the EPSC and consequently LTP can easily be increased in our devices by enlarging the $d_c$ of the semiconducting film, which in turn is determined by the amount of underetching of the top electrode in Fig. 2c. An extreme case is shown in the inset to Fig. 5b where only PDPP is sandwiched between the electrodes (for fabrication details see Supplementary Fig. 11). In these devices the ESPC is increased by a factor of almost 3000 after the last spike and was still increased by a factor of 50 after 10 min. The larger channel area and therefore the larger volume for potential bulk gating results in an increased memory formation compared to smaller channel areas and thus enhanced LTP compared to devices with shorter $d_c$.

Besides synaptic plasticity, also the minimum energy required for a switching operation is a critical factor for possible integration of memristors into complex neuronal networks. Given the large on-state current densities, high on-off ratios and low operation voltages of our devices, we can tune the currents and also switching energies across a wide range depending on the choice of applied voltages. The minimal switching energies we have achieved so far are in the $10^{-13} - 10^{-14}$ J range (see Supplementary Fig. 12), where $V_{DS}$ = 100 µV and $V_{GS}$ = -0.4 to -1.2 V were used. Such low switching energies are already below what is currently used in CMOS neuromorphic devices, and only one magnitude larger than the 10 fJ per event used in the brain[31]. Furthermore, the here obtained switching energies are only a factor of 100 larger compared to the best reported switching energies that have been obtained in core-sheath nanowires[45].

In summary we have reported the fabrication process and electrical characterization of a novel nanoscopic vertical transistor (VOFET) architecture with an electrode separation of 40 nm and a minimal footprint of 2 x 80 x 80 nm² (neglecting the gate contact). Utilizing the high capacitances of liquid electrolytes, our VOFETs can continuously sustain current densities of above 2 MA/cm² at -0.3 V bias with on-off ratios up to $10^8$ and ultra-high transconductances of up to 5 kS/m. Furthermore, we showed that the novel electrolyte-gated VOFET structures can be also operated at low driving voltages down to 10 µV and be utilized as versatile memristive elements: depending on operation voltage and exact transistor layout, the relative susceptibility of the memristive element to STP and LTP can be tuned. Additionally, switching events that require only 10 – 100 fJ / event were realized. We expect the novel VOFET structure to be also interesting for other fields of research such as to assess the vertical mobility of semiconductors for use in solar cells or – given the high current densities in the MA/cm² regime – possibly also for electrically driven lasing[46,47]. A further next step is to increase the switching speed of our in this respect unoptimized devices from 1kHz (Supplementary Fig. 13) to the already demonstrated 10 MHz switching speeds of other electrolyte gated transistors[53].

**Materials and Methods**

*Electrode fabrication:*

Before further treatment, all substrates were cleaned for 10 minutes each in an ultrasonic bath in acetone and isopropyl alcohol and blow-dried with nitrogen. The electrodes were patterned by electron beam lithography (e-line system, Raith). For the bottom electrodes a 250 nm thick layer of PMMA 950K (Allresist) was spin-coated followed by a 150 °C softbake of 3 min. After electron beam exposure, the resist was developed for 100 s in a 1:3 methylisobutylketon : isopropyl alcohol mixture and 1 - 3 nm chromium (evaporated at 0.3 Å/s), 30 nm gold (evaporated at 1 Å/s) and 0.5-1 nm titanium (evaporated at 0.1 Å/s) were evaporated in an electron beam evaporator. Chromium or titanium are used for better adhesion between gold and $SiO_2$ whereby chromium is only evaporated below the bottom electrode since it is resistant to hydrofluoric (HF) acid. The lift-off was performed in acetone with low-power sonication. For the top electrodes a 500 nm layer of PMMA 950K (Allresist) was spin-coated and patterned by electron beam lithography as described above. After development, 35 - 50 nm of $SiO_2$ was sputtered with a power of 50 W and an argon pressure of $p = 2 \times 10^{-2}$ mbar. 1 nm titanium and 90 nm gold were evaporated in the same fashion as for the bottom electrode. Titanium is used between the electrodes and can be etched with HF. The sputtered $SiO_2$ was etched with 1% hydrofluoric acid ($d_c$ = 80 – 120 nm).

*Organic semiconductor and gate preparation:*

PDPP (P3HT) was dissolved in 1,3-dichlorobenzene (Sigma-Aldrich) (15 mg/ml) at 80 °C (70 °C for P3HT) and stirred for at least 12h (4 h for P3HT). For planar devices a diluted semiconductor solution was doctor bladed (Zehntner ZAA 2300, ZUA 2000, ACC225). For the vertical devices the semiconductor solution was spin-coated on the substrate for 40 s (1000 rpm) and followed by a bake of 2 min at 80°C. By directional reactive ion etching (Oxford Plasmalab 100 ICP RIE) with an oxygen plasma the semiconductor is removed everywhere except below the top contacts which serve as etching masks (20 sccm, 20 mbar, 10 W, 180 – 210 s). The liquid electrolyte [EMIM][TFSI] was dropped with a syringe such that the crossing point of bottom and top electrode are covered. Before electrical measurements the samples were stored in a vacuum oven (50 °C, 10 mbar, > 12 h) to bake out moisture residues from the electrolyte.

*Electrical characterization:*

Measurements were performed unless noted otherwise at ambient condition with a conventional point probe station and two Source Meters (Keithley 2450) as source drain and gate bias, respectively. For connecting the gate, the needle was immersed in the liquid electrolyte. Synaptic plasticity measurements were performed with a Keysight 33500B waveform generator and for switching response measurements additionally a PeakTech 1325 USB oscilloscope was used. Vacuum measurements were performed in a Lakeshore CRX-VF probe station.

## Acknowledgements

We would like to thank BASF SE for supplying the organic semiconductors and liquid electrolytes.

## Funding

The authors acknowledge partial support by the "Solar Technologies go Hybrid" (SolTech) initiative, the Center for Nanoscience (CeNS) and the Nanosystems Initiative Munich (NIM).

## Author contributions

J.L. and R.T.W conceived the project. J.L. prepared the VOFET samples, conducted the measurements and data analysis. F.d.G prepared the lateral transistor samples, conducted the measurements and data analysis. All authors discussed the data. J.L and R.T.W. wrote the manuscript with the input from all authors. R.T.W. supervised the project.

## Competing interests

The authors declare no competing interests.

## Materials & Correspondence

Correspondence and requests for materials should be addressed to R.T.W.

## Data availability

All data is available in the main text or the supplementary materials.

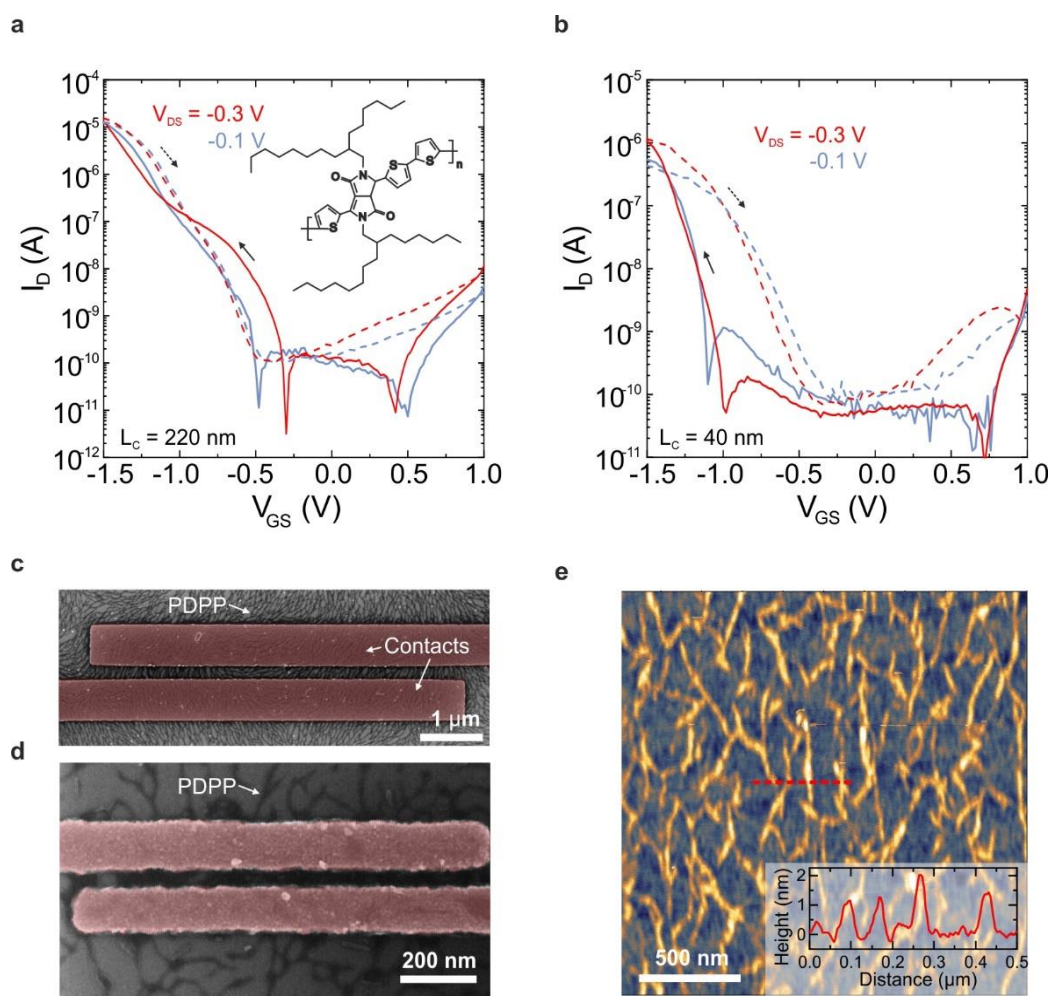

*Fig. 1. Planar, electrolyte-gated nanoscopic PDPP OFETs.* Transfer characteristics of two transistors with (**a**) $L_c$ of 220 nm (Inset: molecular structure of the PDPP used) and (**b**) 40 nm respectively and corresponding SEM images (**c**) and (**d**) (measured at 32 mV/s under vacuum). **e,** Atomic force microscopy topography image of a $SiO_2$ substrate with sub-monolayer PDPP-coverage (top) allowing to measure the height of individual polymer nanofibers (bottom).

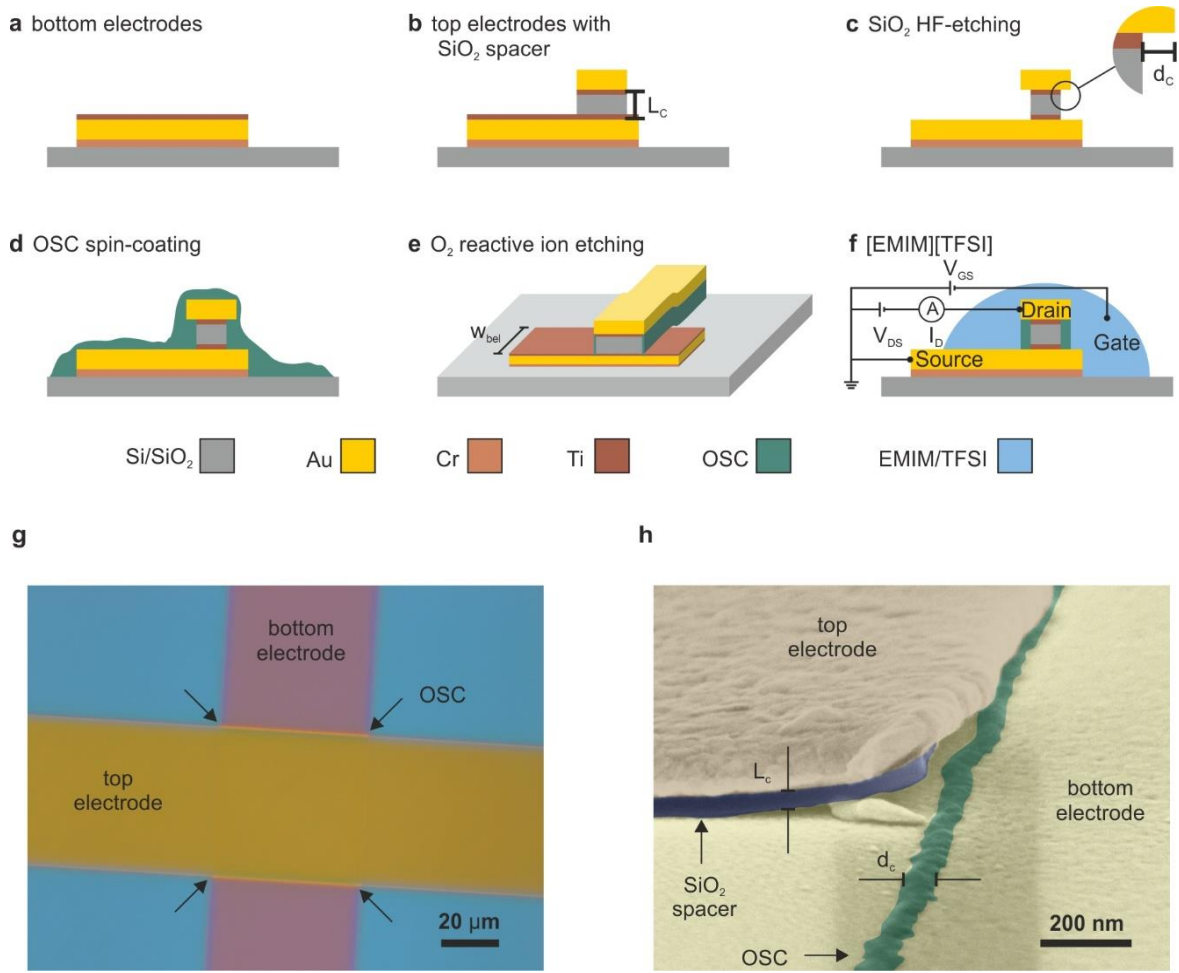

*Fig. 2. Device fabrication process of electrolyte gated VOFETs. Patterning of bottom (a) and top (b) electrode by e-beam lithography. c, HF underetching of the top electrode. d, Spin-coating of the organic semiconductor solution and (e) subsequent directional oxygen RIE. f, Fully finished transistor. g, Polarization microscopy image of a finished VOFET without electrolyte gate. The PDPP appears bright between the electrodes (labelled OSC in the figure). h, Colored cross-sectional SEM image of a VOFET with the two gold electrodes (yellow), SiO$_2$ spacer (blue) and PDPP (green).*

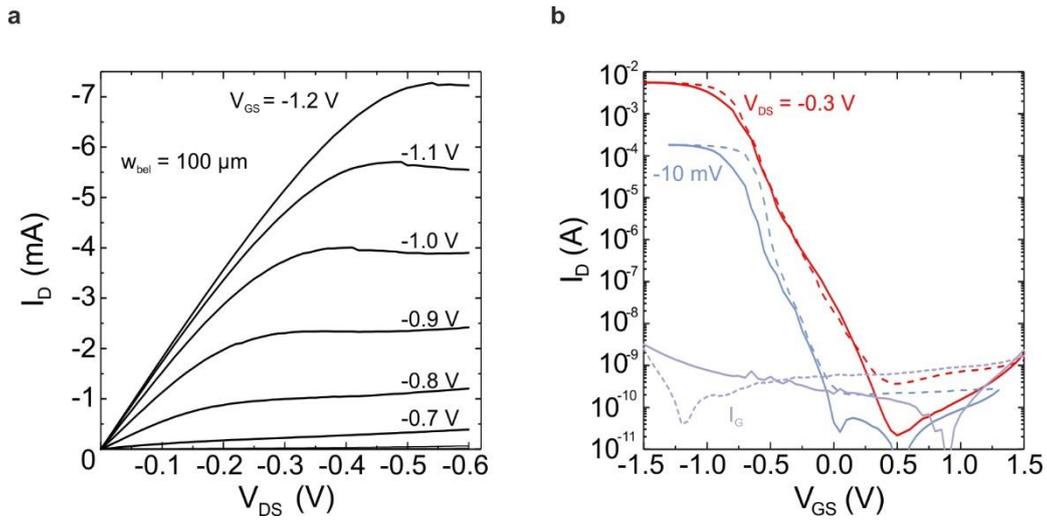

*Fig. 3. Electrical characteristics of electrolyte gated PDPP VOFET measured in ambient atmosphere. **a,** Output characteristics measured at 16 mV/s. **b,** Transfer characteristics with corresponding gate current $I_G$ measured at 80 mV/s. At $V_{DS}$ = -0.3 V, a high on-off ratio of $10^8$ and an on-state current density of 34.9 kA/cm$^2$ are achieved. The size of the device: $w_{bel}$ = 100 µm, $d_c$ = 80 nm, $L_c$ = 40 nm.*

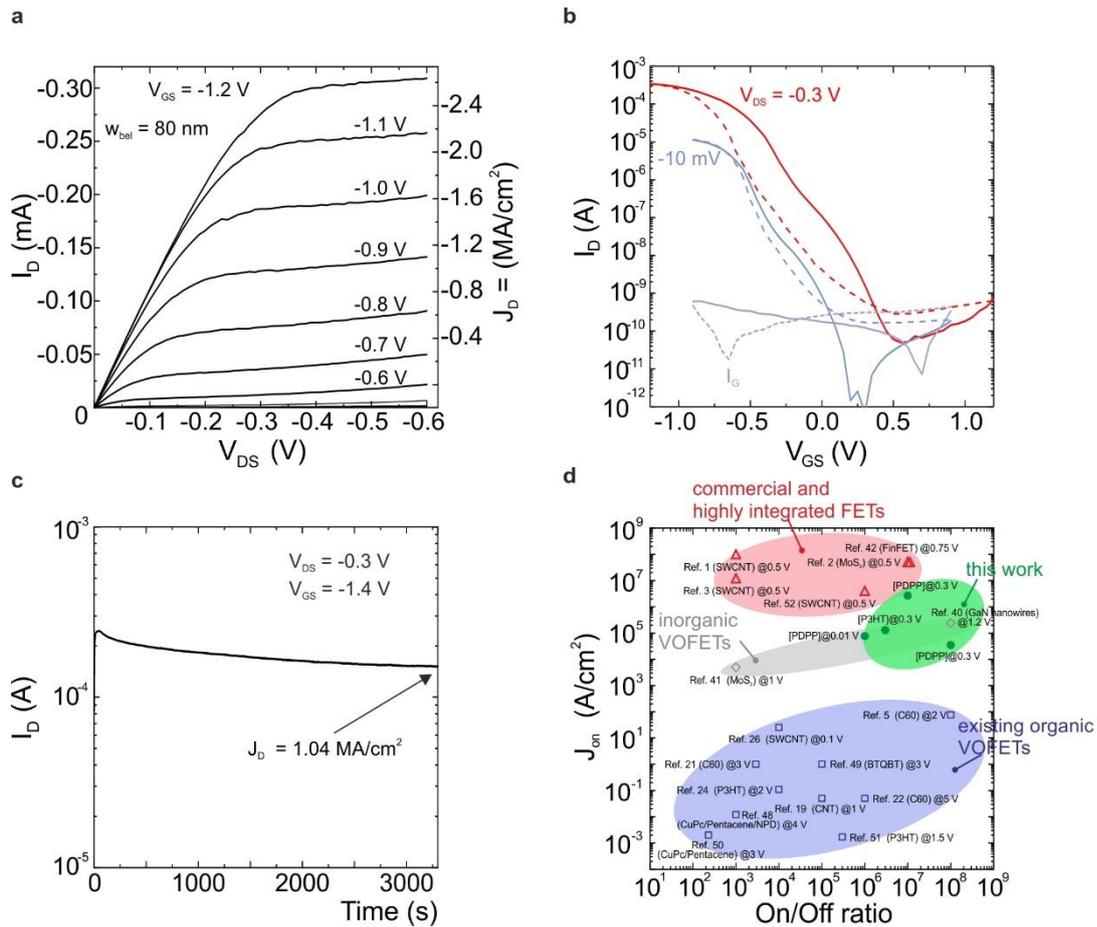

Fig. 4. Nanoscopic electrolyte-gated PDPP VOFET measured in ambient atmosphere and comparison to state of-the-art transistors. a, Output characteristics measured at 16 mV/s and (b) transfer characteristics measured at 50 mV/s of a nanoscopic VOFET with a current density of 2.7 MA/cm². The device dimensions are: $d_c$ = 80 nm, $w_{bel}$ = 80 nm, $L_c$ = 40 nm. c, Continuous operation for 55 min above 1 MA/cm² ($w_{bel}$ = 90 nm, $d_c$ = 80 nm, $L_c$ = 40 nm).      d, Comparison of on-state current densities and on-off ratios for different vertical and planar transistors. The respective operation voltages $V_{DS}$ are indicated as well.

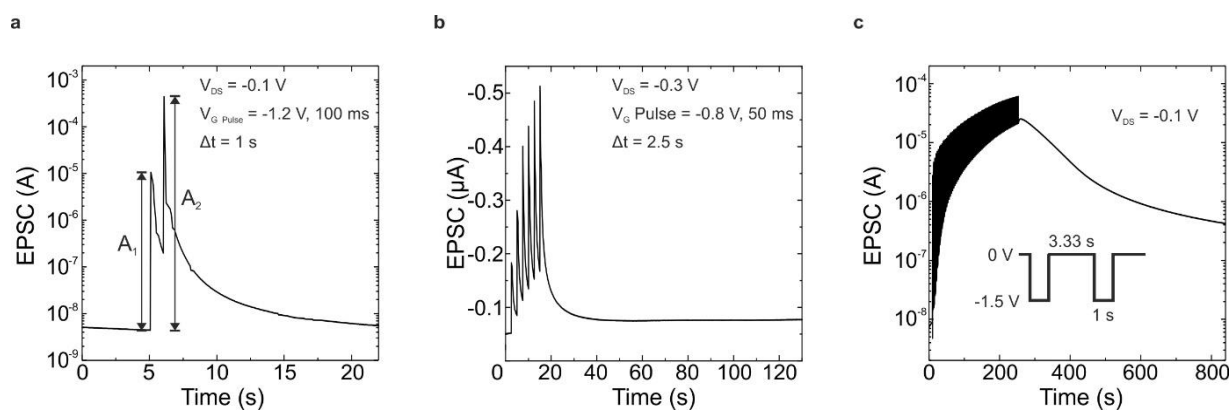

*Fig. 5. Short and long term synaptic plasticity of electrolyte gated PDPP VOFETs measured in ambient atmosphere. **a,** EPSC triggered by two pre-synaptic spikes (-1.2 V, 100 ms) at an inter-spike interval of 1 s ($d_c$ = 100 nm, $w_{bel}$ = 500 µm, $L_c$ = 40 nm). **b,** EPSC versus time simulated by six gate pulses (-0.8 V, 50 ms) and an interval of 2.5 s. **c,** EPSC triggered by 73 pulses (-1.5 V, 1 s) at an inter-spike interval of 3.33 s for an electrolyte gated VOFET without SiO$_2$ spacer and only PDPP between the two electrodes (see Fig. S11).*